\documentclass[a4paper,oneside]{article}       
\RequirePackage[T1]{fontenc}

\RequirePackage{graphicx}
\RequirePackage{mathptmx}      
\RequirePackage{flushend}
\RequirePackage[colorlinks,citecolor=blue,urlcolor=blue,linkcolor=blue]{hyperref}

\usepackage{array}
\usepackage{multirow}
\usepackage{rotating}
\usepackage{hyperref}
\usepackage{float}
\usepackage[margin=2.5cm]{geometry}
\usepackage{rotating}
\usepackage{authblk}

\newcommand\T{\rule{0pt}{2.6ex}}
\newcommand\B{\rule[-1.2ex]{0pt}{0pt}}

\begin{document}

\title{Characterisation of the muon beams for the Muon Ionisation Cooling Experiment}

\author[ ]{The MICE Collaboration}
\author[16]{D.~Adams}
\author[23,b]{D.~Adey}
\author[19,c]{A.~Alekou}
\author[19,d]{M.~Apollonio}
\author[12]{R.~Asfandiyarov}
\author[23]{J.~Back}
\author[19]{G.~Barber}
\author[16]{P.~Barclay}
\author[5]{A.~de Bari}
% ...



% ...
\affil[5]{Sezione INFN Pavia and Dipartimento di Fisica Nucleare e Teorica, Pavia, Italy}
% ...
\affil[12]{DPNC, Section de Physique, Universit\'e de Gen\`eve, Geneva, Switzerland}
% ...
\affil[16]{STFC Rutherford Appleton Laboratory, Harwell Oxford, Didcot, UK}
% ...
\affil[19]{Department of Physics, Blackett Laboratory, Imperial College London, London, UK}
% ...
\affil[23]{Department of Physics, University of Warwick, Coventry, UK}
% ...
\affil[b]{Now at Fermilab, Batavia, IL, USA}
\affil[c]{Now at CERN, Geneva, Switzerland}
\affil[d]{Now at Diamond Light Source, Harwell Science and Innovation Campus, Didcot, Oxfordshire, UK}
% ...

\date{Received: date / Accepted: date -- Entered later}

\maketitle

\begin{abstract}
A novel single-particle technique to measure emittance has been developed and used to characterise seventeen different muon beams for the Muon Ionisation Cooling Experiment (MICE). The muon beams, whose mean momenta vary from 171 to 281\,MeV/$c$, have emittances of  approximately 1.2--2.3\,$\pi$\,mm-rad horizontally and 0.6--1.0\,$\pi$\,mm-rad vertically, a horizontal dispersion of 90--190\,mm and momentum spreads of about 25\,MeV/$c$. There is reasonable agreement between the measured parameters of the beams and the results of simulations. The beams are found to meet the requirements of MICE.

\end{abstract}

\section{Introduction\label{sec:Introduction}}

A future high-energy Neutrino Factory or Muon Collider will require an intense source of muons. The large volume of phase space occupied by muons at production must be reduced before they are accelerated and stored. The short muon lifetime prohibits the use of conventional cooling techniques; another technique must be developed to maximise the muon flux delivered to a storage ring.

Ionisation cooling is the only practical approach. A muon passing through a low-$Z$ material loses energy by ionisation, reducing its transverse and longitudinal momenta. The longitudinal momentum is restored by accelerating cavities, with the net effect of reducing
the divergence of the beam and thus the transverse phase space the beam occupies.

The muon beams at the front-end of a Neutrino Factory or Muon Collider will be similar.  They are expected to have a very large transverse normalised emittance of $\varepsilon_{N} \approx$ 12--20\,$\pi$\,mm-rad and momentum spreads of 20\,MeV/$c$ or more about a central momentum of 200\,MeV/$c$. The transverse emittance must be reduced to 2- 5\,$\pi$\,mm-rad (depending on the subsequent acceleration scheme) for a Neutrino Factory \cite{NFReference,FS2A,nf,nfISS}.  Further transverse and longitudinal cooling is required for a Muon Collider. Emittances of $0.4\,\pi$\,mm-rad and $1\,\pi$\,mm-rad are desired in the transverse and longitudinal planes respectively, where the latter is achieved by emittance exchange \cite{higgsFactory}.

The Muon Ionisation Cooling Experiment (MICE) will be the first experiment to demonstrate the practicality of muon ionisation cooling. This paper describes measurements of the muon beams that will be used by MICE.

\section{The MICE Experiment\label{sec:The-Muon-Ionisation}}

\begin{figure*}[t]
\centering
\includegraphics[scale=0.5]{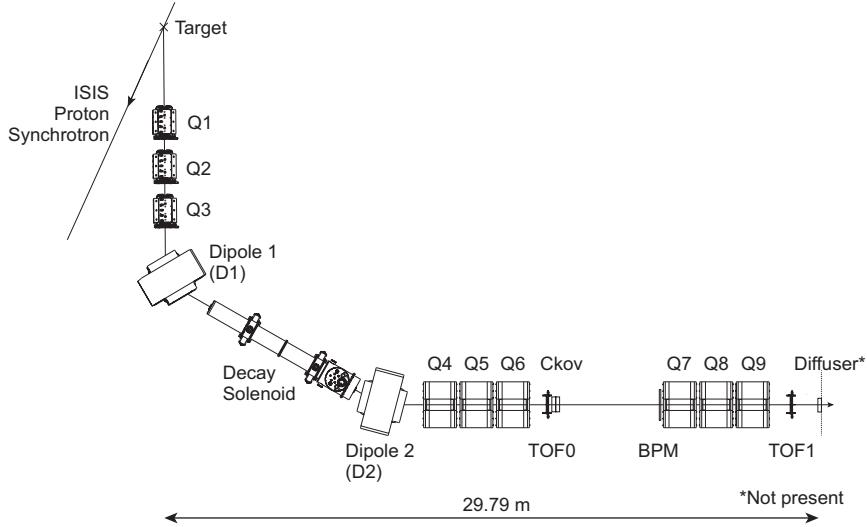}
\caption{MICE upstream beam line.
\label{fig:MICE-upstream-beamline}}
\end{figure*}

MICE will measure the ionisation cooling efficiency of one ``Super Focus-Focus'' (SFOFO) lattice cell~\cite{SFOFO} based on the cooling-channel design of Neutrino Factory Feasibility Study 2 \cite{NFReference}.  A detailed description of the cooling cell is contained in \cite{MICE_TRD}. Since ionisation cooling depends on momentum (via the dependence of energy loss and multiple scattering in materials), the MICE experiment has been designed to measure the performance of the cell for beams of 140 to 240\,MeV/$c$ with large momentum spreads; liquid hydrogen and other low-$Z$ absorber materials will be studied.

The expected reduction in emittance ($\approx$10\% using liquid hydrogen) is too small to be measured conventionally, where methods typically attain 10\% precision.  MICE will therefore make single-particle measurements using scintillating fibre trackers~\cite{TrackerPaper} inside  superconducting solenoids (the ``spectrometer solenoids'') at each end of the cooling cell. Cherenkov detectors and time-of-flight (TOF) detectors provide upstream particle identification; the TOFs will also allow the muons to be timed with respect to the RF phase. A pre-shower detector  and a muon ranger  will provide  particle identification downstream of the cooling section.

\subsection{MICE beam requirements\label{sub:MICE-beam-requirements}}

For a realistic demonstration of cooling the beams used should closely resemble those expected at the front end of a Neutrino Factory, \emph{i.e.}, they should have a large momentum spread and a large normalised emittance. The emittance must be variable to allow the equilibrium emittance---which depends on the absorber material and the optics of the channel---to be measured. 
 
The MICE beam line has been designed to produce beams of three different emittances at each of three central momenta. These beams are named by the convention ``$\left(\varepsilon_{N},p_{z}\right)$'' according to their normalised transverse emittance at the entrance to the cooling section and longitudinal momentum at the centre of the first absorber. The nominal (RMS) input emittances are $\varepsilon_{N} =3$, 6 and $10\,\pi$\,mm-rad; the central momenta are 140, 200 and 240\,MeV/$c$. The baseline beam configuration is $\left(\varepsilon_N,p_{z}\right)$ = ($6\,\pi$\,mm-rad, 200\,MeV/$c$).

The beams of different emittances  will be generated by means of a ``diffuser'', which allows a variable thickness of high-$Z$ material to be inserted into the beam at the entrance to the upstream spectrometer solenoid. Scattering increases the emittance of the beam to the desired values and, as a consequence of the energy lost in the material, beams with a higher emittance downstream of the diffuser must have a higher  momentum upstream. An important requirement is that the muon beam downstream of the diffuser is correctly matched in the spectrometer solenoid.

\subsection{MICE muon beam line\label{sub:MICE-beamline}}

The new muon beam line for MICE (at the ISIS proton synchrotron,  Rutherford Appleton Laboratory) is shown in Figure~\ref{fig:MICE-upstream-beamline} and described at length in~\cite{BeamlinePaper}. A titanium target \cite{target} samples the proton beam, creating pions that are captured by the upstream quadrupoles (Q1--3) and momentum-selected by the first dipole (D1). The beam is transported to the Decay Solenoid, which focuses the pions and captures the decay muons.

The second dipole (D2) can be tuned to select muons emitted backwards in the pion rest frame to obtain a high purity muon beam. The final transport is through two large-aperture quadrupole triplets, Q4--6 and Q7--9, that focus the beam onto the diffuser. Each of the three quadrupole triplets is arranged to focus-defocus-focus in the horizontal plane; the beam line can be operated in either polarity. The optics of this section are determined by the desired emittance of the beam in the cooling channel and the requirement of matching into the spectrometer solenoid.

A time-of-flight station (TOF0) and two aerogel Cherenkov detectors are located just after the Q4--6 triplet; a second time-of-flight station (TOF1) is located after the final triplet (Q7--9). A low-mass scintillating-fibre beam-position monitor (BPM) is located close to Q7. For the $\mu^+$ beams, a variable thickness polyethylene absorber is introduced upstream of D2 to reduce the flux of protons incident on TOF0.

The TOF detectors are described in~\cite{BeamlinePaper,TOFref}. Each station consists of two perpendicular ($x,y$) planes of 25.4\,mm thick scintillator slabs. Each end of each slab is coupled to a fast photomultiplier and subsequent electronics \cite{TOFref3}. The measured timing resolutions are $\sigma_{t}=55$\,ps and $\sigma_{t}=53$\,ps at TOF0 and TOF1 respectively \cite{anothertofref}. The differences in the arrival times of light at each end of the slabs are used to obtain transverse position measurements with resolutions of $\sigma_{x}=9.8$\,mm at TOF0 and $\sigma_{x}=11.4$\,mm at TOF1~\cite{MarksThesis}.

\subsection{Beam line design \label{sub:BL-design}}


The initial design of the baseline $\left(\varepsilon_N,p_{z}\right)$ = ($6\,\pi$\,mm-rad, 200\,MeV/$c$) $\mu^{+}$ beam was made using the TURTLE beam transport code \cite{turtle} assuming a 1\,cm thick lead diffuser. The design was then optimised with G4beamline~\cite{G4BL}, with matching conditions in the upstream spectrometer solenoid of $\alpha_{x}=\alpha_{y}=0$ and  $\beta_{x}=\beta_{y}=333$\,mm~\cite{BeamlinePaper}.  The baseline beam design does not compensate for horizontal dispersion introduced at D2.  The remaining  $\left(\varepsilon_N,p_{z}\right)$ beam settings were obtained by scaling the magnet currents of the baseline case according to the the local muon momentum, accounting for the energy loss of muons in the beam line material, \emph{i.e.} scaled by a factor $p_{\mathrm{new}}/p_{\mathrm{base}}$.  
Hence, the beam line will transport 18 different beams to the cooling channel with $\varepsilon_{N}=3,6,10\,\pi$\,mm-rad and $p_{z}=140,200,240$\,MeV/$c$, after the diffuser, in two beam polarities.

The ``re-scaled'' beam line settings will transport muons to the cooling channel with the desired momenta but are not necessarily matched in the first spectrometer as scattering in the diffuser changes the optical parameters. Because the diffuser is thin, the beta function will decrease by the same ratio as the emittance is increased and therefore the final optics and diffuser thicknesses cannot be determined until the inherent emittances of the input beams are known.




The beam line was commissioned in MICE ``Step I'' in 2010--2011. Data were taken for eight positive and nine negative beam settings to verify the beam line design and determine the characteristics of the beams, in particular their momentum distributions, emittances and dispersions.  The result of the commissioning is presented below.

\section{Characterising the MICE beams\label{sec:Characterising-the-MICE-beam}}


Emittance is the area occupied by a charged particle beam, in two, four, or six-dimensional trace-space, given by $\varepsilon = \sqrt{\det\Sigma}$ where $\Sigma$ is the covariance matrix.  In two dimensions, 
\begin{displaymath}
	\Sigma = \left( \begin{array}{cc}
			\sigma_{xx}     &        \sigma_{xx'}    \\
			\sigma_{x'x}   &      \sigma_{x'x'}
			\end{array} \right)
			\equiv
			\left( \begin{array}{cc}
			\varepsilon \beta    &       - \varepsilon \alpha    \\
			- \varepsilon \alpha   &      \varepsilon \gamma
			\end{array} \right),
\end{displaymath}
where, for example, $\sigma_{xx}=\overline{x}\,\overline{x} - \overline{xx}$ and $\overline{x}$ denotes the mean.  The covariance matrix can also be expressed in terms of the Twiss parameters $\alpha, \beta, \gamma$, and $\varepsilon$ giving a full parameterisation of the beam.

Several different methods exist for measuring the emittance of beams~\cite{McDonald}.  Commonly, beam profile monitors are used to measure the RMS beam size, $\sigma_{x}$, at several positions.  At least three profile monitors are required to determine the three elements of the covariance matrix and hence the emittance of the beam; the transfer matrices between the profile monitors must be known.  These methods do not require individual particles to be tracked but are ultimately limited by the spatial resolution of the detectors and the intensity of the beam.

By contrast, the MICE muon beam is large in spatial extent and its intensity is low compared to conventional primary beams.  The emittance and optical parameters of such a beam can be measured if either the trace space co-ordinates, $(x,x'), (y,y')$ of individual particles can be measured at a single plane or, as in the new method described here, the spatial co-ordinates of individual particles are measured at two detectors and the transfer matrix between the detectors is known.

In the later Steps of MICE the beam emittance will be measured by a scintillating fibre tracker inside a 4T solenoid.  This detector was not present during Step I and the new method was developed to characterise the beam using only the two TOF detectors. The relative times and $(x,y)$ positions of single particles are measured in the two TOF stations and muons are selected by broad time-of-flight cuts. Each muon is tracked through the Q7--9 quadrupole triplet, determining the trace space angles $x'$ and $y'$ at each plane.  Simultaneously, the muon momenta is measured by time-of-flight, which is important as the beam has a large momentum spread and the transfer matrix between the two detectors depends strongly on momentum.  The covariance matrix of the beam is then obtained from a large sample of muons so measured.  The method is described briefly below; its detailed development is given in~\cite{MarksThesis}.

\begin{figure}

\centering
\includegraphics[scale=0.6]{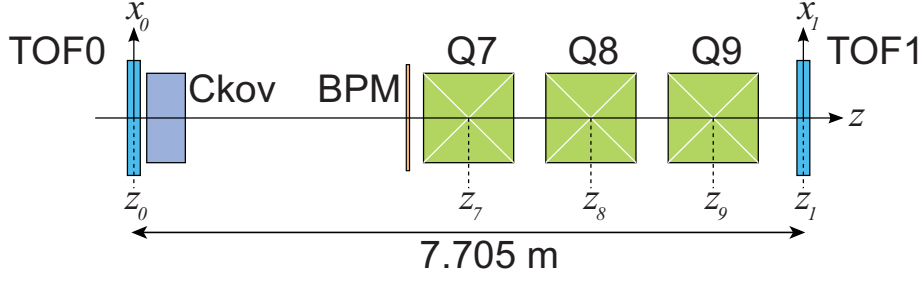}

\caption{The time-of-flight system and beam line section used
to characterise the beam.\label{fig:The-MICE-time-of-flight-system}}
\end{figure}

Figure~\ref{fig:The-MICE-time-of-flight-system} shows the section of beam line used for the measurements. The pole tip radius of the quadrupoles is 176\,mm. The two TOF detectors have active areas of 400\,mm $\times$ 400\,mm and 420\,mm $\times$ 420\,mm, respectively, and were separated by 7.705\,m during the 2010 commissioning; their combined time resolution of 76\,ps allows the momenta of muons to be determined with a resolution of $\sigma_p = 3.7$\,MeV/$c$ for $p_z = 230$\,MeV/$c$.

\subsection{Measurement technique\label{sub:Trace-space-reconstruction}}

The measurement algorithm proceeds iteratively. An initial estimate of $p_{z}$ is made by assuming that a muon travels along the $z$-axis between the two TOF counters. This estimate is used to determine the $x$ and $y$ transfer matrices, $M_{x}(p_{z})$ and $M_{y}(p_{z})$, between TOF0 and TOF1. Once the transfer matrices are known, the trace-space vectors $(x_{0},x_{0}^{\prime})$ and $(y_{0},y_{0}')$, and $(x_{1},x_{1}')$ and $(y_{1},y_{1}')$, at TOF0 and TOF1 respectively, are obtained from the position measurements $(x_{0},y_{0})$ and $(x_{1},y_{1})$ by a rearrangement of the transport equations:
\begin{equation}
\left(\begin{array}{c}
x_{1}\\
x_{1}^{\prime}
\end{array}\right)=M_{x}\left(\begin{array}{c}
x_{0}\\
x_{0}^{\prime}
\end{array}\right),
\nonumber
\end{equation}
\begin{equation}
\left(\begin{array}{c}
y_{1}\\
y_{1}^{\prime}
\end{array}\right)=M_{y}\left(\begin{array}{c}
y_{0}\\
y_{0}^{\prime}
\end{array}\right)\,.
\label{eq:Trace-space-transport-equation}
\end{equation}
Explicitly
\begin{equation}
\left(\begin{array}{c}
x_{0}^{\prime}\\
x_{1}^{\prime}
\end{array}\right)=\frac{1}{M_{12}}\left(\begin{array}{cc}
-M_{11} & 1\\
-1 & M_{22}
\end{array}\right)\left(\begin{array}{c}
x_{0}\\
x_{1}
\end{array}\right),
\label{eq:Trace-Space_TransferMatrix}
\end{equation}
where $M_{ij}$ are the (momentum dependent) elements of $M_{x}$, and similarly for $(y_{0}',y_{1}')$. The estimates of $(x_{0},x_{0}')$, $(y_{0},y_{0}')$, and $p_{z}$ are used to track the muon between TOF0 and TOF1 and obtain an improved estimate
of the trajectory and a correction, $\Delta s$, to the path length. To ensure convergence to a stable solution, only half the predicted $\Delta s$ was applied before recalculating the momentum from the time-of-flight; the procedure was repeated ten times for each muon although a convergent solution was found after typically five iterations. 
Finally, a small correction ($\approx$ 1.5\,MeV/$c$) is applied to account for energy loss in the material, including air, between the TOF counters.

\begin{figure}[tb]
\centering
\includegraphics[scale=0.6]{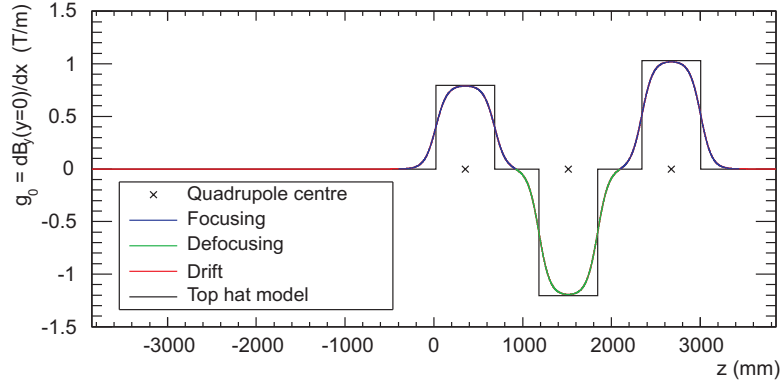}
\caption{
Quadrupole gradients for the ($6\,\pi$\,mm-rad, 200\,MeV/$c$) baseline muon beam (colour online).}
\label{fig:gradients}
\end{figure}

In order to obtain the transfer matrices, the focusing gradients of quadrupoles Q7--9 were determined by fitting the results of an OPERA~\cite{OPERARef} field model of the quadrupole with two hyperbolic tangent functions~\cite{MarksThesis}. Figure~\ref{fig:gradients} shows the focusing gradients of the Q7--9 triplet. The quadrupoles are thick and their fields overlap substantially. A more computationally efficient, and sufficiently accurate, ``top-hat'' model of the magnets was used to obtain $\Delta s$~\cite{MarksThesis}.

\begin{figure}[h!]
\includegraphics[scale=0.44]{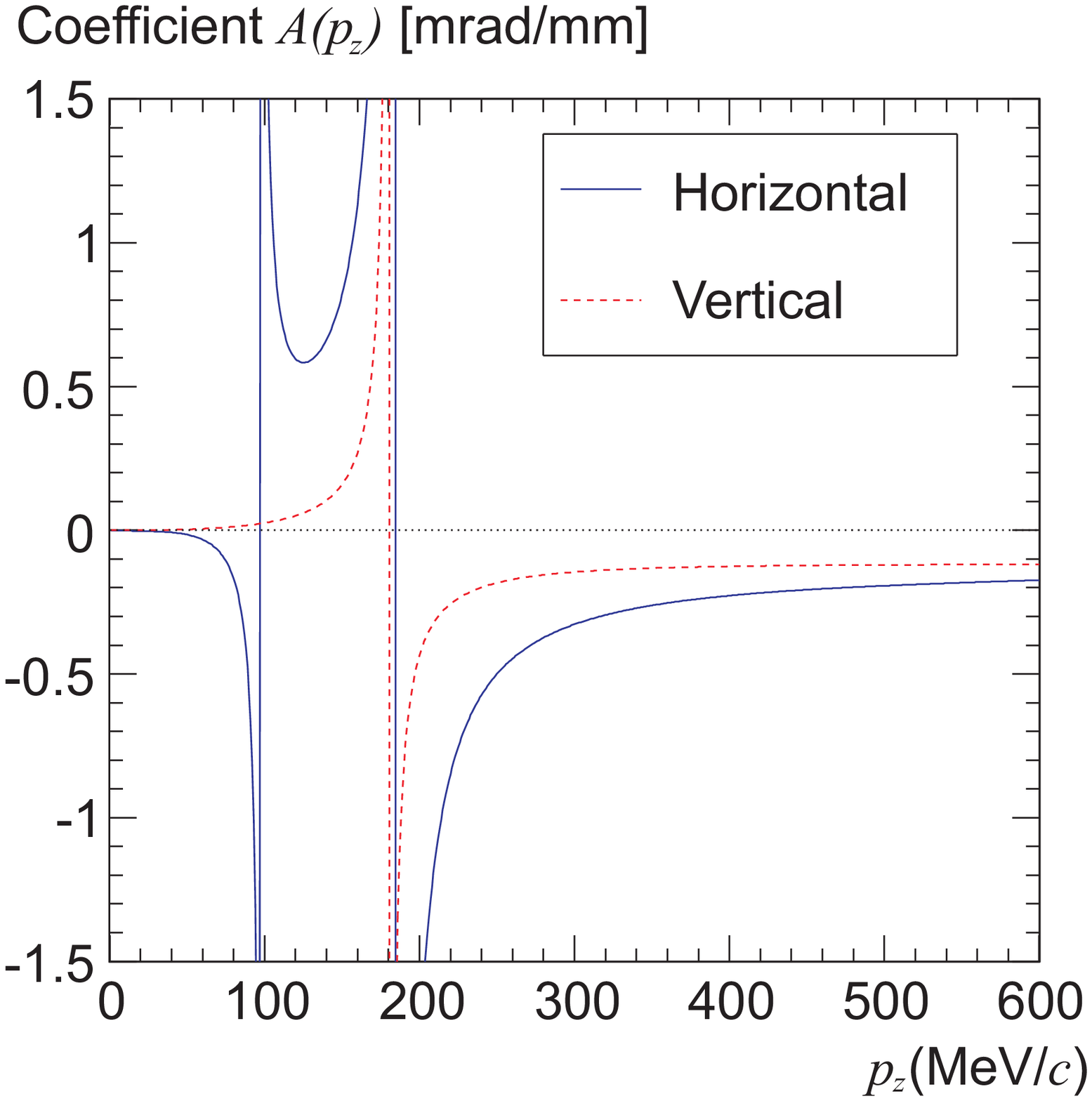}~
\includegraphics[scale=0.44]{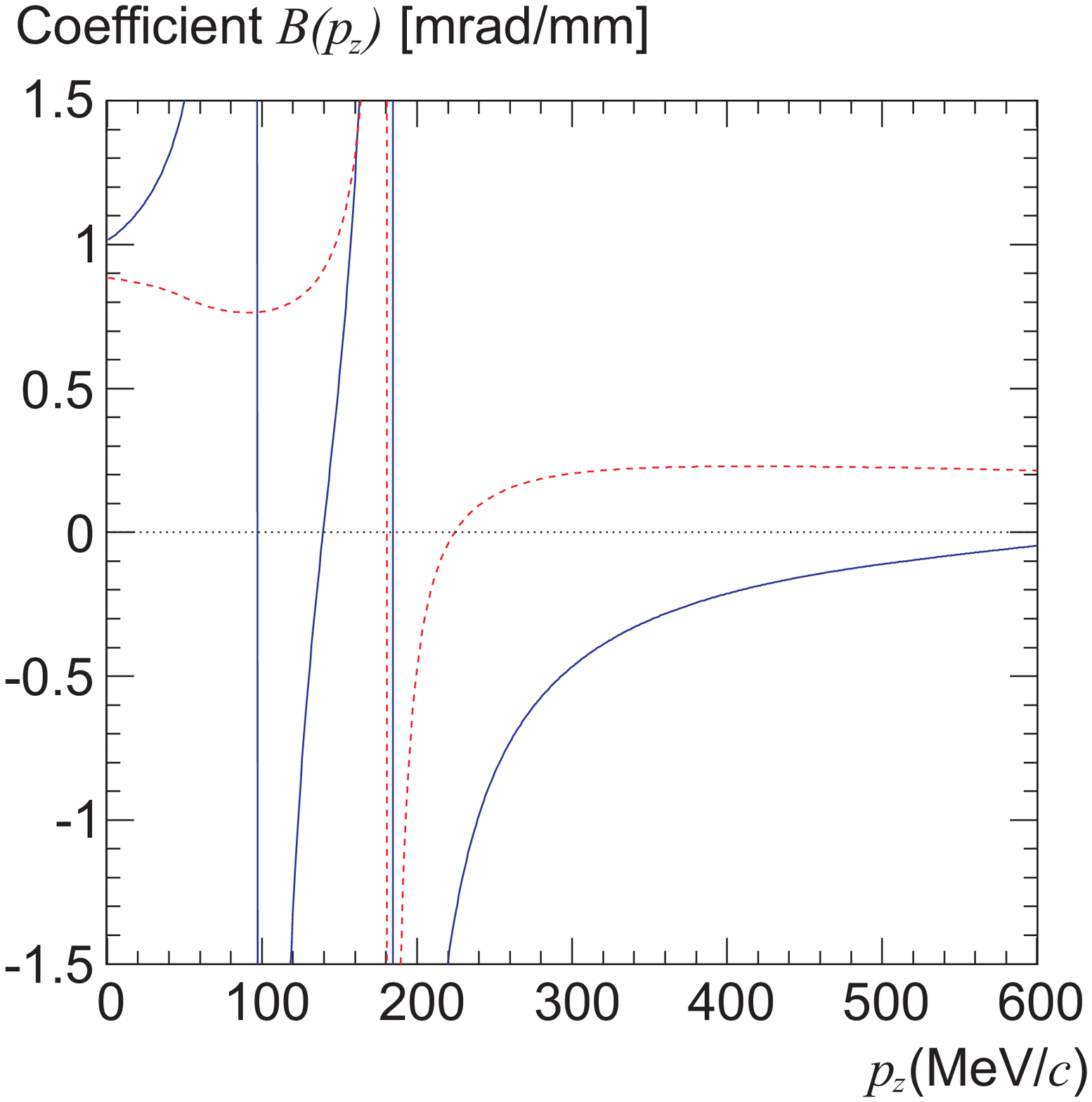}
\caption{
The reconstruction coefficients $A(p_{z})$ (top) and $B(p_{z})$ (bottom) for the (6,\,200) baseline muon beam. The solid (blue) lines are for $x'$ (horizontal); the dashed (red) lines are for $y'$ (vertical).}
\label{fig:ABcoeffs}
\end{figure}

Equation~\ref{eq:Trace-Space_TransferMatrix} for $x_{1}'$, which is used to determine the horizontal emittance at TOF1, can be expressed as 
\begin{displaymath}
x_{1}' = A(p_{z}) x_{0} + B(p_{z}) x_{1}
\end{displaymath}
and {\it mutatis mutandis} for $y'$. The coefficients $A(p_{z})$ and $B(p_{z})$ for the baseline (6,\,200) beam, with mean $p_{z} \approx 230$\,MeV/$c$, are shown in Figure~\ref{fig:ABcoeffs}. Both $A$ and $B$ are strongly momentum dependent below 200\,MeV/$c$.

The procedure described above enabled the reconstruction of the trace space vectors at both TOF counters as well as the momenta of single muons. The path length correction, which could be as much as 15--20\,mm, was necessary to avoid a systematic underestimate of $p_{z}$ of about 4\,MeV/$c$.

The momentum distributions and the $(x_{1},x_{1}')$ and $(y_{1},y_{1}')$ covariance matrices, $\Sigma_{x,y}$, at the upstream side of TOF1 for each measured beam were obtained from all the muons recorded for that beam. The effective optical parameters and the emittances of each beam were deduced from $\Sigma_{x,y}$ as described in Section~\ref{Sec:EmittanceAndOpticalParameters}. The systematic uncertainty on the measurements is discussed in Section~\ref{Sec:Errors} 

\subsection{Monte Carlo simulations of the MICE beam line}

Monte Carlo simulations were made for six of the 18 possible beam settings to check the beam line design software.  The ($6\,\pi$\,mm-rad, $p_{z}$ = 140, 200, 240\,MeV/$c$) $\mu^+$ and $\mu^-$ beams were simulated in two steps. 
G4beamline was used to track particles from the target as far as TOF0; the G4MICE Monte Carlo \cite{G4MICE} was then used to track muons between TOF0 and TOF1. Both simulations contained descriptions of the material in, and surrounding, the beam line  and magnet models, including the apertures of the quadrupoles Q4--9, using the optics designed for the corresponding beams. The simulations suggest  that the final emittance of the beams before the diffuser is $\approx 1\,\pi$\,mm-rad, partly due to scattering in the material in the beam line but limited by the aperture of the quadrupoles. Dispersion in the horizontal plane due to D2 is expected. 

\subsection{Performance of the reconstruction algorithm\label{sub:Performance-of-reco}}

\begin{figure}[h!]
\centering
\includegraphics[scale=0.44]{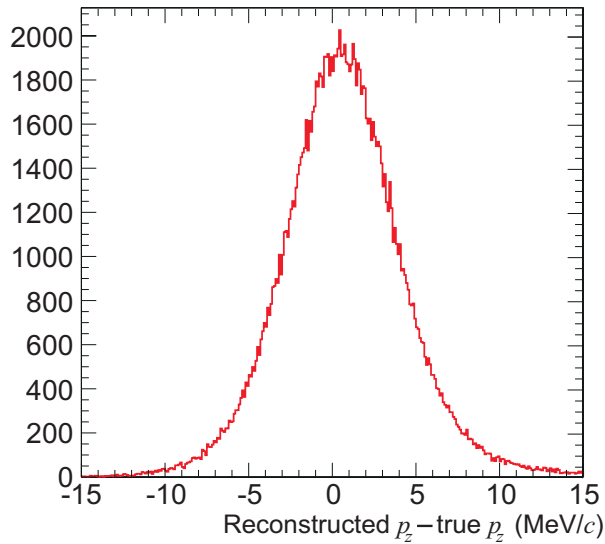}
\caption{
Difference between reconstructed and true momenta for a simulated 200\,MeV/$c$ muon beam.}
\label{fig:MarkFigures_5pt7_5pt10_5p14_5pt_15}
\end{figure}

\begin{figure*}[tbp]
\centering
\includegraphics[scale=0.8]{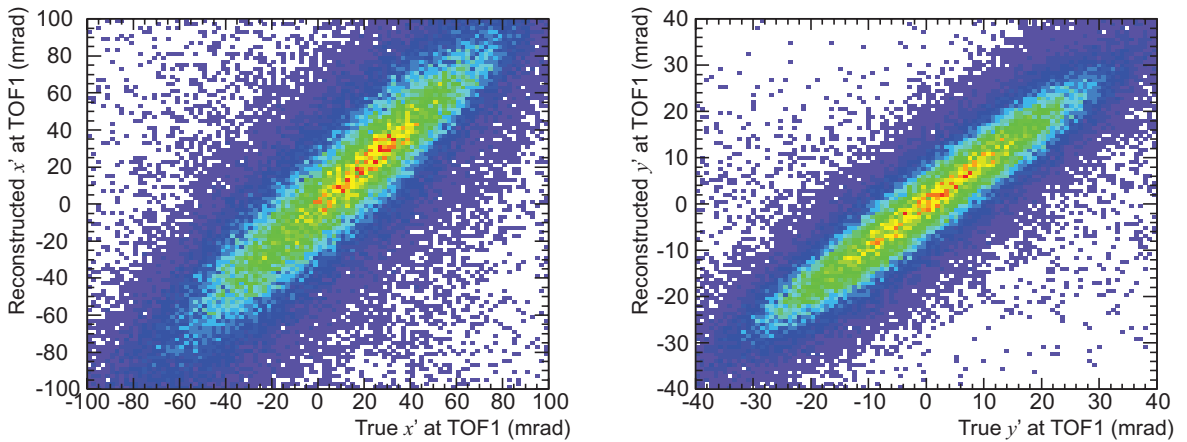}
\caption{Reconstructed trace space angles versus true simulated values.
\label{fig:MarkFigure_5pt18}}
\end{figure*}

The performance of the reconstruction algorithm was determined by smearing the true simulated coordinates of the muons at each TOF plane with the measured time and position resolutions of the TOFs.  The trace-space vectors were reconstructed by the method described in  Section \ref{sub:Trace-space-reconstruction} to obtain a set of simulated reconstructed muons.

Figure~\ref{fig:MarkFigures_5pt7_5pt10_5p14_5pt_15} shows, for a simulated $(6,200)\,\mu^{-}$ beam, the difference between reconstructed and true momenta. The RMS width of the distribution of 3.7\,MeV/$c$ confirms that the momentum resolution is dominated by the timing resolution of the TOF system. It is sufficiently small to measure the large expected widths of approximately 20\,MeV/$c$ of the momentum distributions.

Figure~\ref{fig:MarkFigure_5pt18} shows the agreement between the true and reconstructed angles, $x_{1}^{\prime}$ and $y_{1}^{\prime}$ for the simulated $(6,200)\,\mu^{-}$ beam.  The average angular resolutions, $\sigma_{x_{1}'}$ and $\sigma_{y_{1}'}$, for this beam are approximately 29 and 8\,mrad respectively. They are determined by the position resolution of the TOF counters and multiple scattering, and depend on momentum as $x_{1}'$ and $y_{1}'$ are obtained from position measurements using the momentum-dependent elements of the transfer matrix. The angular resolutions are small but not negligible compared with the expected widths of the $x_{1}'$ and $y_{1}'$ distributions.

\section{Results of the measurements and comparison with simulations}

Data were taken for eight positive and nine negative re-scaled beams that, when used in conjunction with the diffuser, will generate the full range of desired emittances (see Section~\ref{sub:BL-design}); the polarity of the decay solenoid
was kept the same for both positive and negative beams.  Muons in the data  were selected by broad time-of-flight cuts chosen for each nominal beam momentum. 

The simulations used in this analysis suggest that the pion contamination at TOF0 of the $\mu^-$ data is about one percent and less than five percent for the $\mu^+$ sample~\cite{MarksThesis}. Recent measurements \cite{KLpaper} indicate a somewhat smaller pion contamination.

\begin{figure*}[p]
\centering
\includegraphics[scale=0.9]{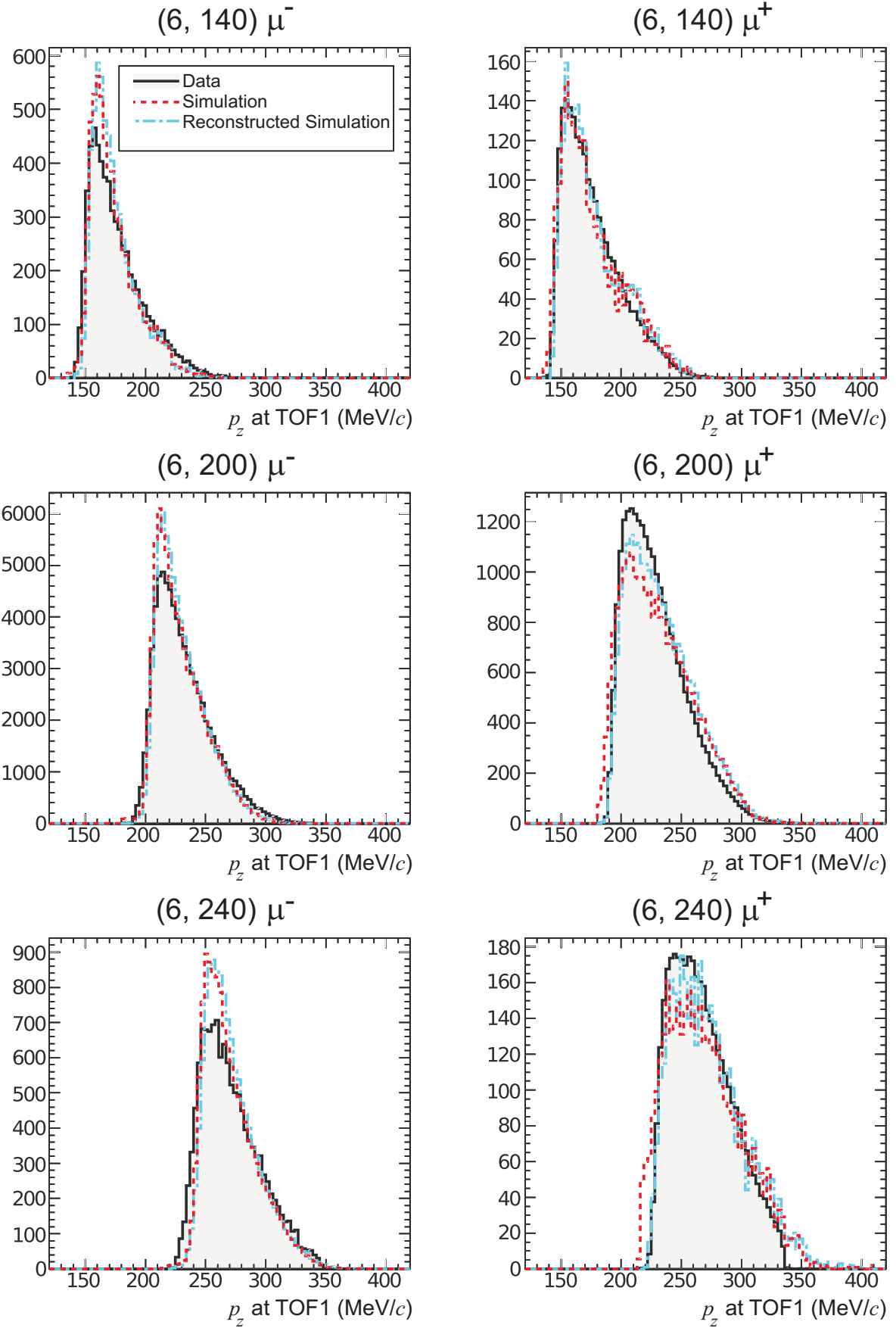}
\caption{Reconstructed longitudinal momentum distributions at TOF1 for six MICE beams compared with simulations. The dotted (red), dash-dotted (blue) and shaded distributions are simulation, reconstructed simulation and data respectively. Distributions are normalised to contain equal numbers of events.}
{\label{fig:six-momentum-distributions}}
\end{figure*}

\subsection{Longitudinal momentum}
\begin{table*}
\centering
\caption{Mean and RMS widths of the longitudinal momentum distributions for six beams compared with the corresponding simulations.
\label{tab:PzInDataAndSimulation}}

\begin{tabular}{cccccc}
\hline
 &  & \multicolumn{2}{c}{Data} & \multicolumn{2}{c}{Simulation}\T\B\\ 
\multicolumn{2}{c}{Beam}  & Mean $p_{z}$ & RMS $p_{z}$ & Mean $p_{z}$ & RMS $p_{z}$ \\ 
 & & {MeV/$c$}& {MeV/$c$}& {MeV/$c$}& {MeV/$c$}\B\\ 
\hline
& $\left(6,140\right)$ &176.4$\pm$2.3 & 22.8$\pm$0.3 & 173.7$\pm$2.1 & 19.5$\pm$0.2\T\\ 
$\mu^-$& $\left(6,200\right)$ & 232.2$\pm$2.5 & 23.6$\pm$0.3 & 229.3$\pm$0.8 & 21.0$\pm$0.1\\ 
& $\left(6,240\right)$ &271.0$\pm$3.7 & 24.5$\pm$0.3 & 270.5$\pm$0.9 & 22.2$\pm$0.1\B\\ 
\hline
 & $\left(6,140\right)$ & 176.5$\pm$2.0 & 24.4$\pm$0.3 & 176.6$\pm$3.7 & 25.5$\pm$0.5\T\\ 
 $\mu^+$& $\left(6,200\right)$ &229.2$\pm$2.4 & 25.9$\pm$0.3 & 230.8$\pm$1.4 & 28.9$\pm$0.2\\ 
 & $\left(6,240\right)$ & 267.7$\pm$2.9 & 25.8$\pm$0.3 & 269.2$\pm$4.2 & 31.3$\pm$0.5\B\\ 

\hline

 \end{tabular} 

 \end{table*}

Figure~\ref{fig:six-momentum-distributions} shows the distributions of $p_{z}$ at TOF1 for six beams compared with the results of the simulations. Overall the measured and simulated distributions agree well in shape and width. The $\mu^+$ beams have a slightly greater momentum spread than the $\mu^-$ beams, due to energy loss fluctuations in the proton absorber. The agreement between the measured and simulated momentum distributions is better for the $\mu^-$ beams than it is for the $\mu^+$ beams. 
 Since the mean momentum is dictated by D2, the agreement between the measured and simulated mean momenta at TOF1 confirms the beam line design.
The mean momenta and the RMS widths of the measured beams are given in Table~\ref{tab:PzInDataAndSimulation}. The systematic error on $p_{z}$ is mainly due to the $\pm35$\,ps calibration uncertainty  on the absolute time of flight value  \cite{MarksThesis} and is estimated to be less than 3\,MeV/$c$ for all momenta below 300\,MeV/$c$.

\subsection{Transverse spatial distributions}
\label{sec:transdis}
\begin{figure*}[tb]
\centering
\includegraphics[scale=0.8]{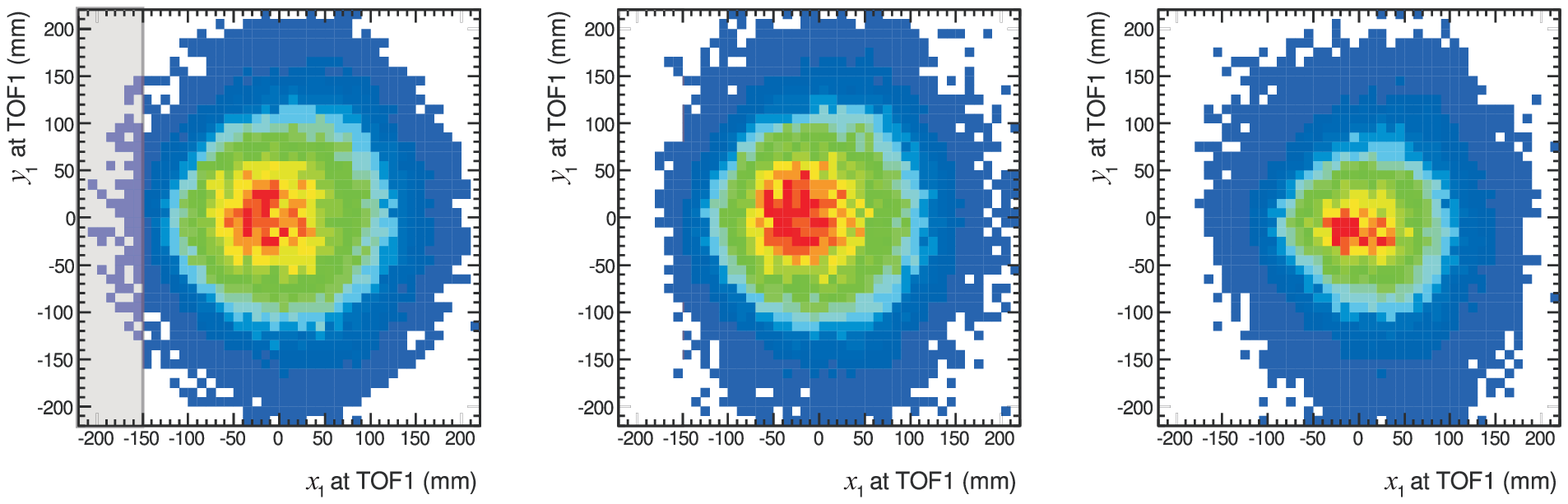}
\caption{
Spatial distributions in the transverse plane at TOF1 for simulation
(left), reconstructed simulation (centre), and data (right) for a (6,\,200)\,$\mu^-$ beam, normalised to the same total contents. Simulated muons in the shaded area cross uncalibrated regions of TOF1  and are excluded from further analysis.
\label{fig:xyplots}}
\end{figure*}

\begin{figure}[tb]
\centering
\begin{tabular}{cc}
\includegraphics[scale=0.42]{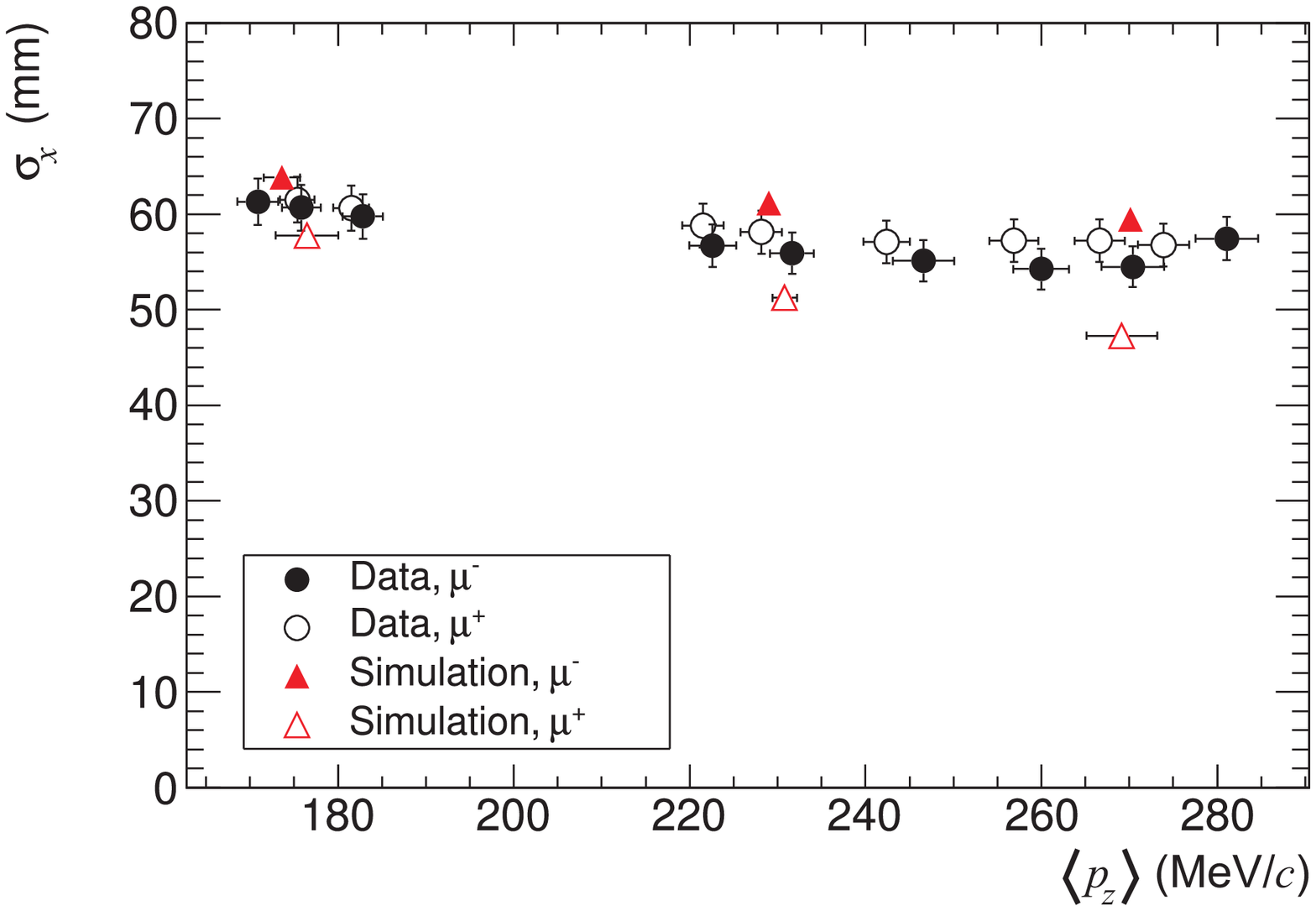}

\\
\includegraphics[scale=0.42]{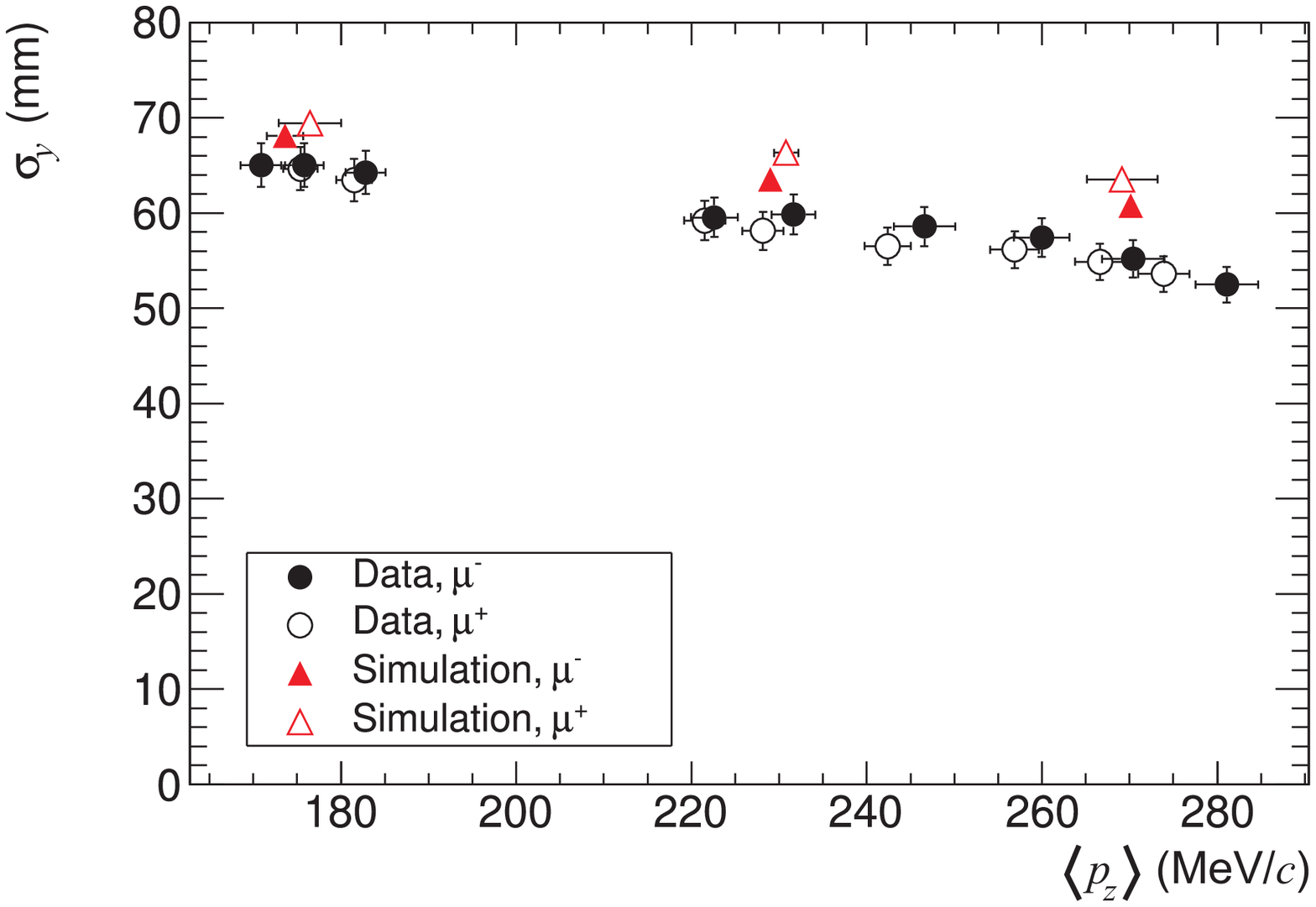}
\end{tabular}
\caption{ Root mean square beam widths, $\sigma_x$, $\sigma_y$, at TOF1 versus $p_z$.
Solid black circles: $\mu^-$ data, open black circles: $\mu^+$ data,
solid red triangles: $\mu^-$ simulation, open red triangles: $\mu^+$ simulation. The nominal  ``$p_{z}=140$''\,MeV/$c$ beams correspond to momenta in the range 170--190\,MeV/$c$, ``$p_{z}=200$'' to 220--250\,MeV/$c$, and ``$p_{z}=240$'' to 250--290\,MeV/$c$.
}
{\label{fig:beamsizes}}
\end{figure}

Figure~\ref{fig:xyplots} shows a comparison of the spatial distributions in the transverse plane at TOF1 for a simulated (6,\,200)\,$\mu^{-}$ beam before and after reconstruction, and data for the same beam. The effect of smearing by the reconstruction procedure is small. Muons crossing the shaded area are excluded from the simulation (and hence the reconstruction) as they pass through uncalibrated regions of TOF1. Since muons must cross both a horizontal and vertical slab of the TOF to be counted, these regions are excluded from the data.  Figure~\ref{fig:beamsizes} shows the RMS horizontal and vertical beam sizes, $\sigma_{x}$ and $\sigma_{y}$, versus mean $p_{z}$ for all the measured beams and the six simulated beams. The sizes of the positive and negative muon beams are very similar both vertically and horizontally. The measured vertical beam size is about 10--20\%  smaller than suggested by the simulations. The horizontal beam size is about 10\% smaller than the $\mu^{-}$ simulations but wider than the $\mu^{+}$ simulations.

\begin{figure*}[p]
\centering
\includegraphics[scale=0.9]{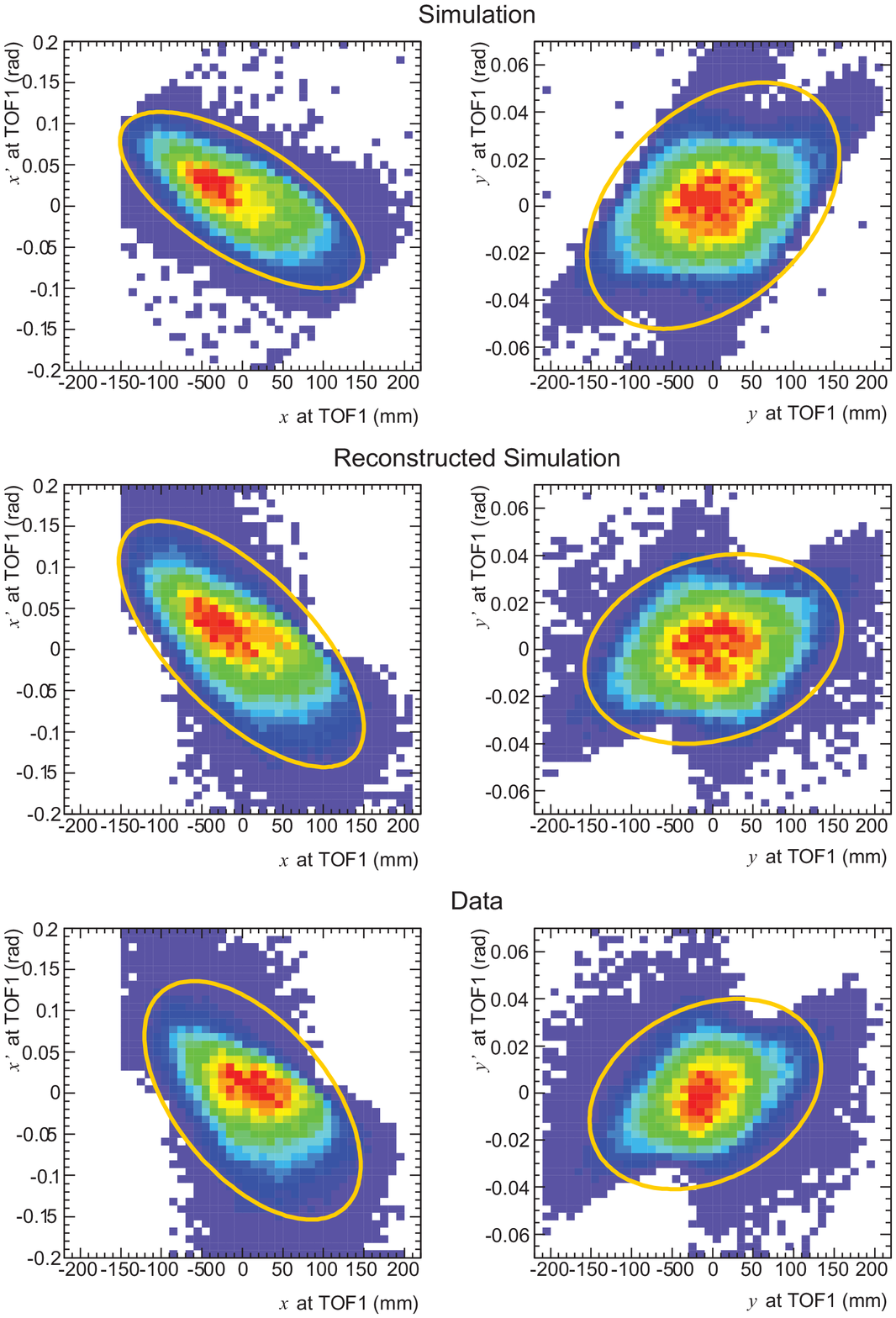}
\caption{Horizontal ($x$) and vertical ($y$) trace space distributions at TOF1 for simulation (top), reconstructed simulation (centre) and data (bottom) for a (6,\,200)\,$\mu^{-}$ beam. The ellipses correspond to $\chi^{2} = 6$ (see text). 
\textbf{
\label{fig:traceSpace}}}
\end{figure*}

\begin{figure*}
\centering
\includegraphics[width=15cm]{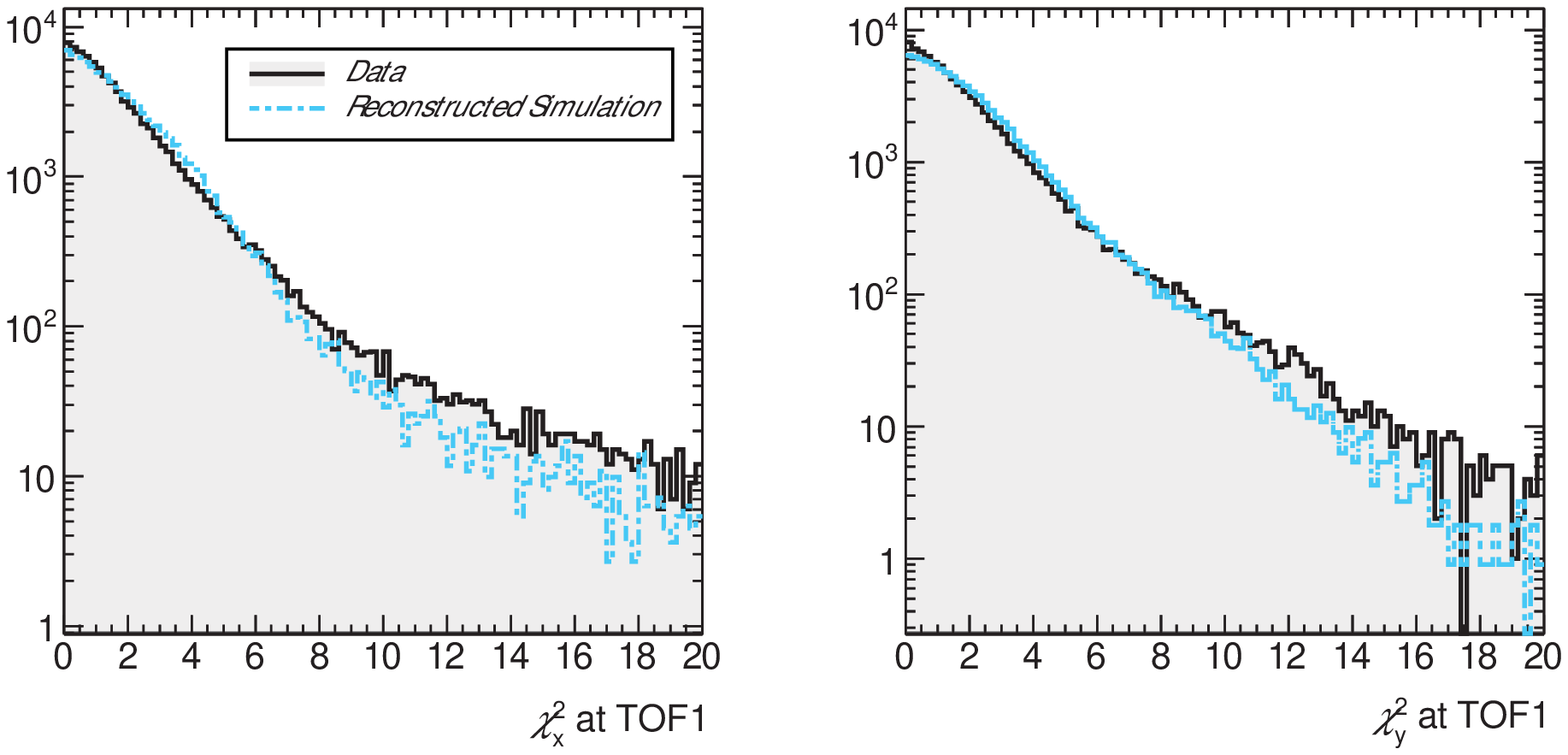}
\caption{
Distributions of $\chi^{2}$ for data (solid, shaded, black) and reconstructed simulation (dash-dot, blue) for the (6,\,200)\,$\mu^{-}$ beam. Left: horizontal $\chi^{2}$; right: vertical $\chi^{2}$ at TOF1.}
{\label{fig:Chisq-distributions}}
\end{figure*}

Figure~\ref{fig:traceSpace} shows the horizontal ($x,x'$) and vertical ($y,y'$) trace-space distributions at TOF1 for the (6,\,200)\,$\mu^{-}$ beam and the same distributions from the simulations with and without smearing due to the reconstruction.   There is very good qualitative agreement between data and reconstructed simulations in both the horizontal and vertical trace spaces. The smearing due to the reconstruction is apparent. The distributions have a dense core and diffuse halo. The boundaries of the distributions reflect the apertures of the quadrupoles, principally Q9, transported to the TOF1 measurement plane downstream, and the size of TOF1. The vertical divergence of the beam is approximately three times smaller than the horizontal divergence.

Figure~\ref{fig:Chisq-distributions} shows the $x$ and $y$ amplitude distributions of muons in the (6,\,200)\,$\mu^-$ beam at TOF1  in terms of $\chi^2_{x,y}$ where
$$ \chi^{2}_{x}  =  [(x-\bar{x}), (x'-\bar{x}')]\Sigma_x^{-1}[(x-\bar{x}), (x'-\bar{x}')]^{T}
= {A_{x}}/{{\varepsilon_{x}}}\,,
$$
$A_{x}$ is the amplitude of a muon in trace-space\footnote{This is sometimes referred to as `single particle emittance'~\cite{Wille}.} and ${\varepsilon_{x}} =\sqrt{\det \Sigma_{x}}$ is the emittance of the ensemble. The distributions of $\chi^{2}$ for the reconstructed simulation are shown for comparison. The initial exponential behaviour of the distribution suggests that the beam has a quasi-Gaussian core up to $\chi^{2} \approx 6$ and a non-Gaussian tail. The high amplitude tails of the distributions are slightly larger for the data than for the simulation. 

\subsection{Determination of emittances and effective optical parameters}\label{Sec:EmittanceAndOpticalParameters} 

The optical parameters and emittances of each beam were determined from the covariance matrices~\cite{Rosenzweig} as 
\begin{eqnarray}
\varepsilon_{x} & = & \sqrt{\det \Sigma_{x}}, \nonumber \\
\beta_{x} & = & \frac{\Sigma_{11}} {\varepsilon_{x}}, \nonumber \\
\alpha_{x} & = & -\frac{\Sigma_{12}} {\varepsilon_{x}},\nonumber 
\end{eqnarray}
and similarly for $y$.  Each of the beams, however, has a large momentum spread and  $\alpha$ and $\beta$ vary with momentum. 
The parameters determined from the measurements are therefore effective parameters which describe the distributions in trace-space. 

The reconstructed covariance matrices at TOF1 will differ from the true covariance matrices because of the finite spatial and angular resolution of the reconstruction, and multiple scattering in the air between the TOFs (which cannot be included in the simple transfer matrices used). The finite resolution leads to a small increase in the apparent emittance of the beams; scattering will lead to an underestimate of the emittances.

A small correction was made for the effects of resolution and scattering by subtracting a ``resolution'' matrix from each measured covariance matrix. The resolution matrices were estimated from the simulations by taking the difference between the covariance matrices of the reconstructed and true simulated beams. These resolution matrices were subtracted from the measured covariance matrices to obtain corrected, measured covariance matrices, \emph{i.e.},
\begin{eqnarray}
\Sigma_{\rm Corrected} & = & \Sigma_{\rm Measured} - \Sigma_{\rm Resolution} \nonumber \\ 
                       & = & \Sigma_{\rm Measured} - (\Sigma_{\rm Reco-sim} - \Sigma_{\rm True-sim})\,. \nonumber  
\end{eqnarray}
Since simulations were made for only the six (6\,$\pi$\,mm-rad, $p_{z}$) beams, the resolution matrices estimated for these beams have been used to correct the measured covariance matrices for other beams at the same nominal momentum.  As variances are very sensitive to outliers, muons in the high amplitude tails of the $(x,x')$ and $(y,y')$ distributions were excluded by requiring $\chi^2_{x,y} < 6$ before the corrected covariance matrices were calculated. The ellipses on Figure~\ref{fig:traceSpace} show the areas of the distributions included by this cut. 

\begin{figure}[tb]
\centering
\includegraphics[scale=0.44]{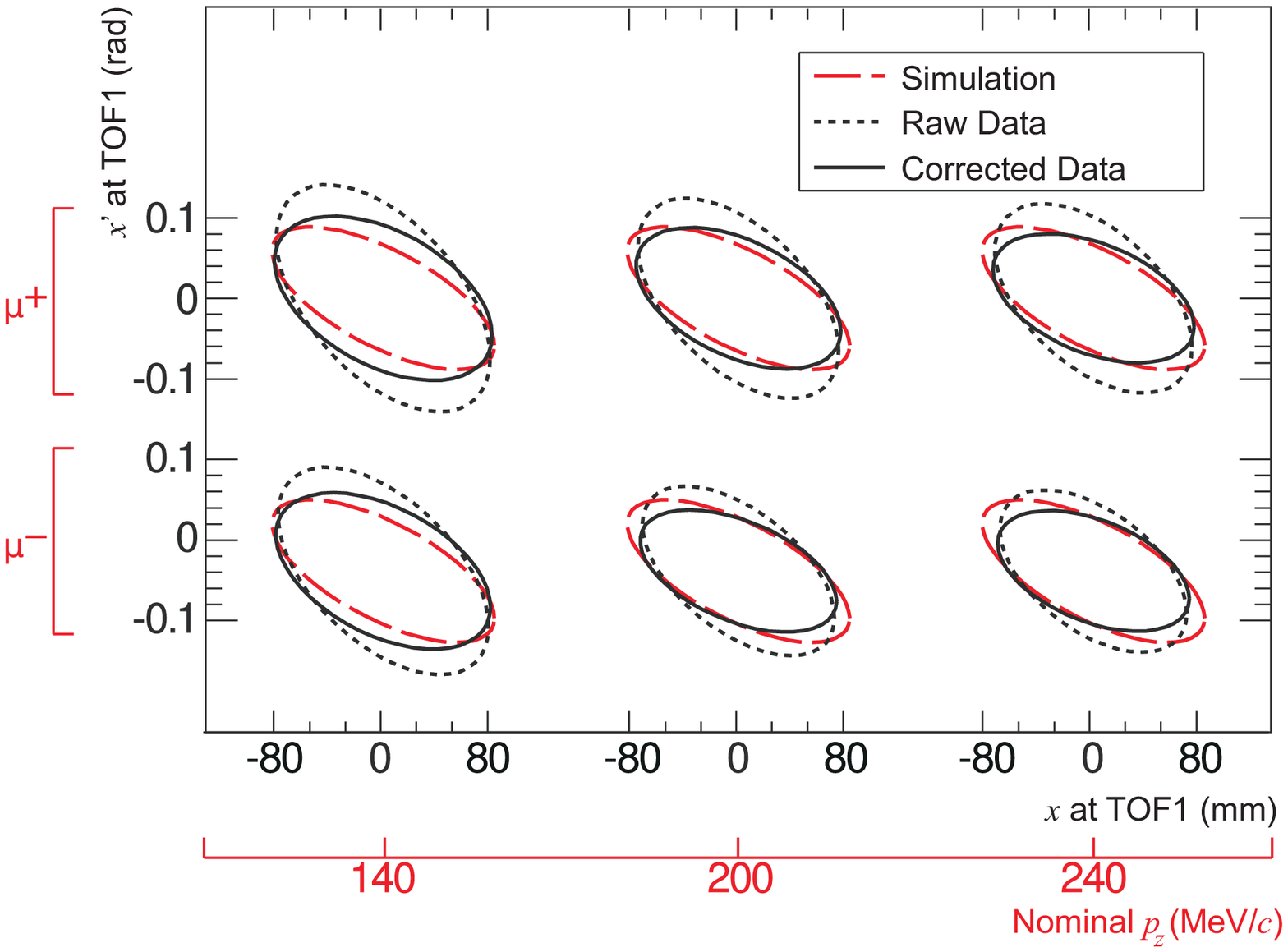}
\caption{RMS emittance ellipses in ($x,x'$) trace-space for data without correction for the measurement resolution (black dotted line),
corrected data (black solid line) and true simulation (red dashed line). 
\label{fig:resellipses}}
\end{figure}

Figure~\ref{fig:resellipses} shows, for the six beams for which simulations were made, the horizontal ($x, x'$) RMS emittance ellipses for the uncorrected data, the data after correction for resolution and the true simulation. The effect of the resolution correction is to reduce the apparent emittance of the beams and to rotate the ellipses into better alignment with the true simulation. 
\begin{figure}[t,b,p]
\begin{centering}
\includegraphics[scale=0.44]{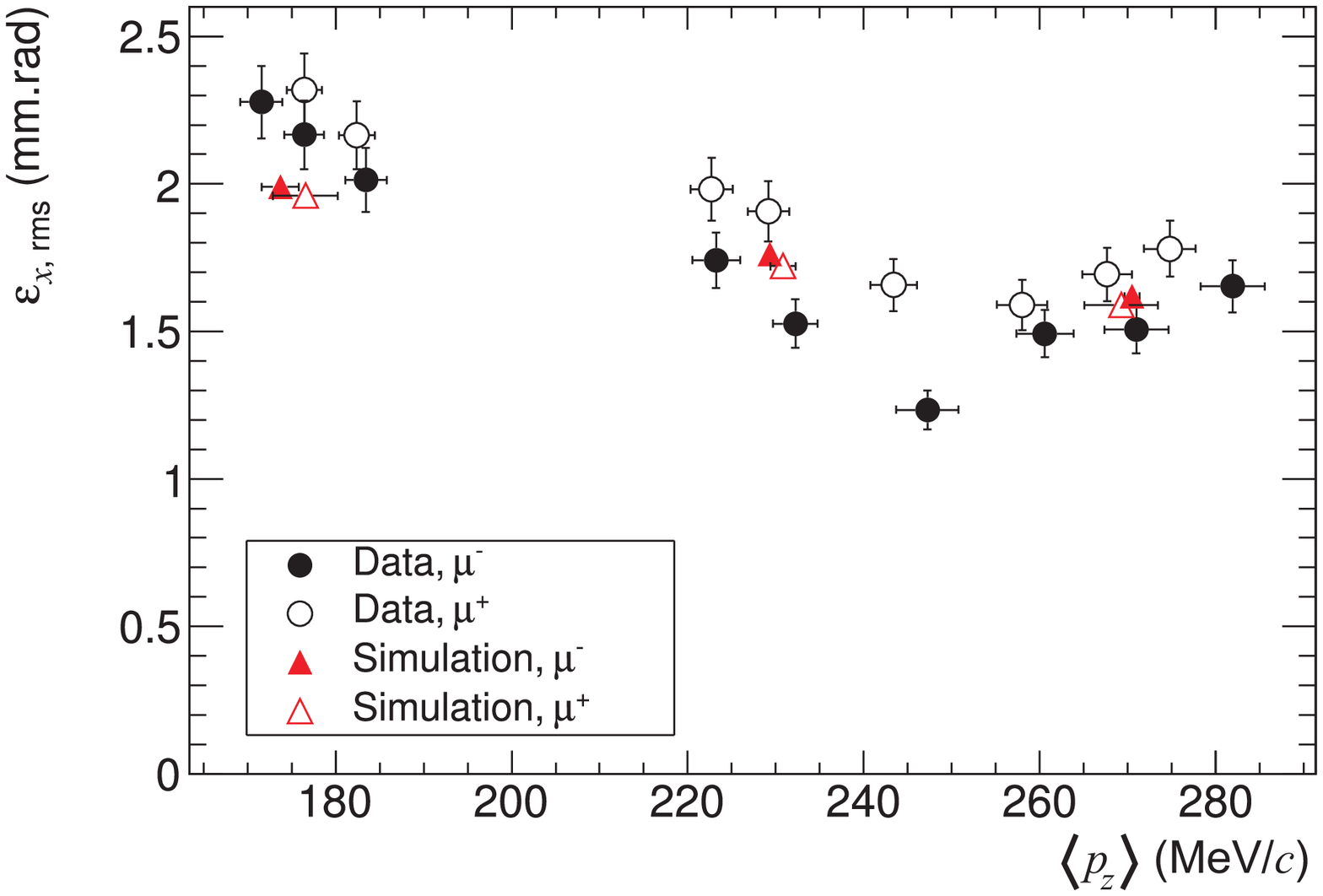}

\end{centering}
\caption{Horizontal emittance after correction for measurement resolution and multiple scattering versus mean $p_{z}$ of the 
seventeen measured beams.  Solid black circles: $\mu^{-}$ data, open black circles: $\mu^{+}$ data, solid red triangles: $\mu^{-}$ simulation, open red triangles: $\mu^{+}$ simulation. The nominal  ``$p_{z}=140$''\,MeV/$c$ beams correspond to momenta in the range 170--190\,MeV/$c$, ``$p_{z}=200$'' to 220--250\,MeV/$c$, and ``$p_{z}=240$'' to 250--290\,MeV/$c$.}\label{fig:Hemit}
\end{figure}

\begin{figure}[t,b,p]
\begin{centering}
\includegraphics[scale=0.44]{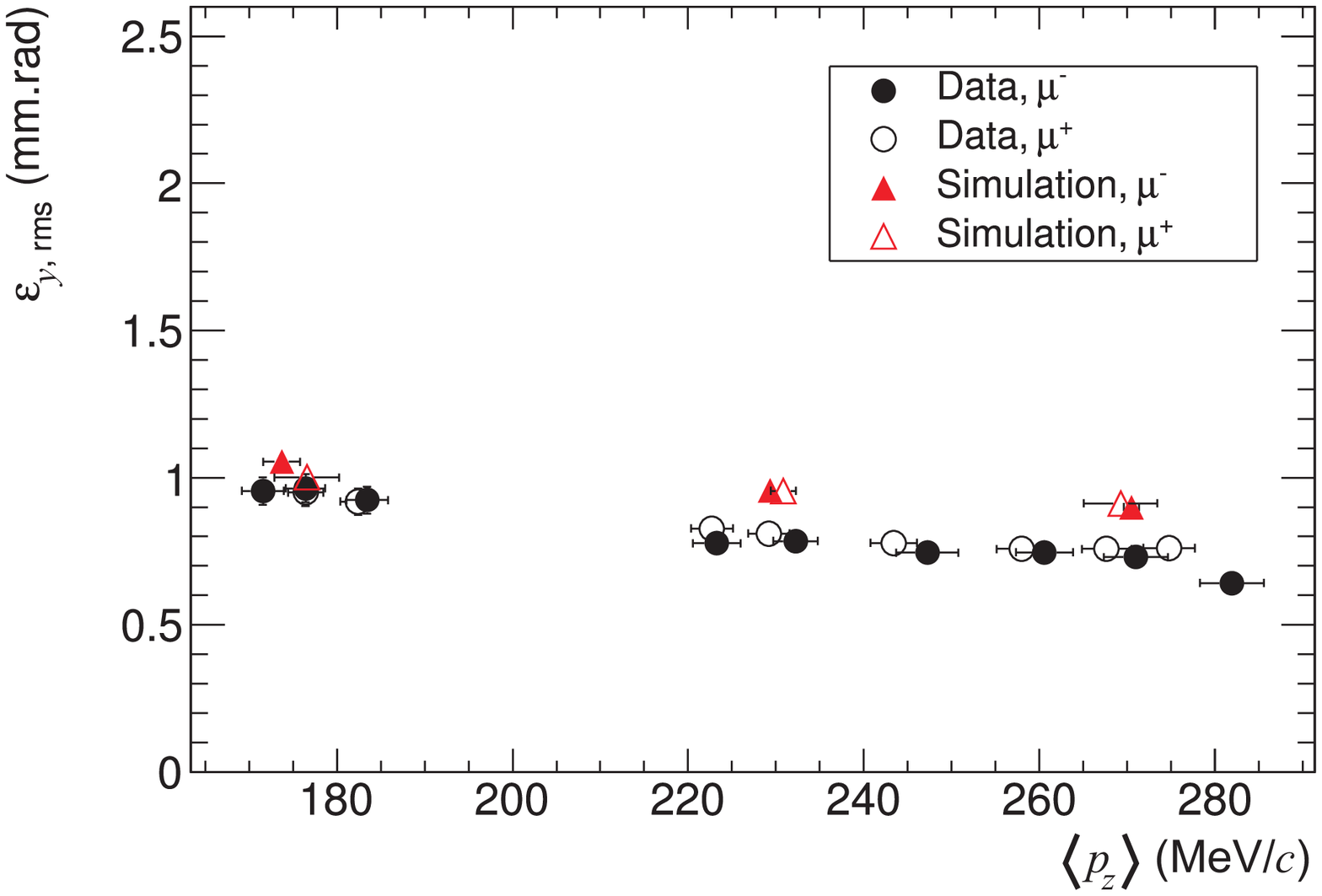}

\end{centering}
\caption{Vertical emittance after correction for measurement resolution and multiple scattering versus mean $p_{z}$ of the 
seventeen measured beams. Solid black circles: $\mu^{-}$ data, open black circles: $\mu^{+}$ data, solid red triangles: $\mu^{-}$ simulation, open red triangles: $\mu^{+}$ simulation. The nominal  ``$p_{z}=140$''\,MeV/$c$ beams correspond to momenta in the range 170--190\,MeV/$c$, ``$p_{z}=200$'' to 220--250\,MeV/$c$, and ``$p_{z}=240$'' to 250--290\,MeV/$c$.
}
\label{fig:Vemit}
\end{figure}

The measured emittances discussed below have not been corrected upward for the $\chi^2_{x,y} < 6$ requirement, which has also been applied to the simulated data, because the long non-Gaussian tail of the amplitude distribution (see Figure~\ref{fig:Chisq-distributions}) is not well-described by the simulations. For a pure Gaussian distribution 5\% of the muons would have $\chi^2 > 6$ and the correction would increase the measured values of emittance by approximately 20\%.  

Figure~\ref{fig:Hemit} shows the measured horizontal emittances, after resolution correction, of all the seventeen beams versus the mean $p_{z}$ of the beam and the true emittances of the six simulated beams. The correction reduces the measured emittances by $0.6\,\pi$\,mm-rad on average; the largest correction is $-0.7\,\pi$\,mm-rad for the (10,140) $\mu^{+}$ beam.  Figure~\ref{fig:Vemit} shows the measured vertical emittances of all the seventeen beams versus the mean $p_{z}$ of the beam and the emittances of the six simulated beams. The correction increases the measured vertical emittances by about 10\%. 
Clipping occurs in the vertical plane as Q4 and Q7 are vertically defocusing.  This collimates the beam, resulting in more uniform emittances compared to the horizontal plane.


\subsection{Systematic uncertainties}\label{Sec:Errors}

The error bars shown on Figures~13 and~14 include both statistical and systematic errors.  Sources of systematic error fall into three broad categories; those that affect the transverse position measurement, momentum reconstruction, and path length corrections.  The largest contribution to the uncertainty on the emittance measurement derives from the effective speed of light in the TOF slabs, which directly determines the measured RMS width of the spatial and angular distributions.   The various sources, summarised in Table~2, were determined by examining the change in the reconstructed emittance and optical parameters when the positions of the TOF detectors and magnet currents were varied in simulation.  

The TOF offsets arise from the uncertainty on their surveyed positions in the beam line.  In each instance, a simulation was produced with one TOF offset by up to 1\,cm in $x,y$ or $z$ and the muon positions recorded.  These positions were input into the reconstruction procedure, which assumes the beam line elements are located as given by survey.  The largest uncertainty occurs when the TOFs are offset in the longitudinal ($z$) direction, which directly affects the momentum measurement by altering the distance $\Delta L$.  

The uncertainty on the quadrupole triplet position in survey was investigated in the same manner as for the TOFs.  However, since this does not affect the distance, $\Delta L$, it has a negligible effect on the momentum calculation and a plays a minor role in the path length correction assigned to a muon.  The currents in the quadrupoles are known to better than 1\%, and the effect of changing these currents was determined.  A change in the quadrupole current by 1\% has a small effect on the reconstructed path length of a muon, when compared to the nominal currents, and is a minor source of uncertainty on the emittance measurement.  The uncertainty on $p_{z}$ has a much larger effect on the transfer matrix used than any scaling due to an uncertainty on the quadrupole currents (\emph{cf.} Figure~\ref{fig:ABcoeffs}).

\subsection{Results}


\begin{table*}[t]
\centering
\caption{Contributions to the errors on the emittance measurements as percentage relative error.\label{tab:errors}}
\begin{tabular}{lcccccccccc}
\hline
\multicolumn{2}{c}{Source} & $\delta \epsilon_{x}$ & $\delta \alpha_{x}$ & $\delta \beta_{x}$ & $\delta \eta_{x}$ & $\delta \eta_{x}'$
						& $\delta \epsilon_{y}$ & $\delta \alpha_{y}$ & $\delta \beta_{y}$ & $\delta p_{z}$\T\B \\ 
\hline
\multirow{2}{*}{TOF1 offsets } & $x$ &0.47         &0.74	& 0.47	&1.39	& 0.69	& 0.014	& 0.05	&$\approx 0$	&$\approx 0$     \T\\
 						& $y$ &    $\approx 0$        &0.01	&$\approx 0$	& 0.29	& 0.17	&0.02	&0.06	& $\approx 0$	&0.71     \\
\multirow{2}{*}{TOF0 offsets} & $x$ & 0.04            & 0.07	&  0.03	& 0.13	& 0.22	& $\approx 0$	&	0.08&  0.01	&   $\approx 0$  \\
 						& $y$& $\approx 0$             & $\approx 0$	&  $\approx 0$	& 0.02	&  $\approx 0$	&$\approx 0$	&	 $\approx 0$& $\approx 0$	&    $\approx 0$ \\
\multicolumn{2}{c}{$\Delta L$  } 	 &   2.10           &	0.32&2.11	&1.69	&3.30	&2.74	&30.17	&	2.78&  0.71   \\
\multicolumn{2}{c}{Q789 currents  } &    0.051          &	$\approx 0$& 0.03	& 0.008	& 0.02	& 0.04	&	0.036 &0.035	&     0.002 \\
\multirow{2}{*}{Q789 offsets  } & $x$&    0.08          &	 0.13& 0.08	& 0.17	&0.99	&	$\approx 0$& 0.08	& 0.01	&  $\approx 0$ \\
						 & $y$ &    $\approx 0$          &$\approx 0$&$\approx 0$	&0.01	&$\approx 0$	&$\approx 0$	&0.01	&$\approx 0$	& $\approx 0$    \\
\multicolumn{2}{c}{Effective $c$ in scintillator  } &       4.87       &	0.05&	5.23&2.22	&	1.59& 4.05 &41.27	&	4.09	&   0.11  \B\\

\hline
\multicolumn{2}{c}{Total (\%)} & 5.32           &  0.82	& 	5.66 & 3.14	& 3.87	&4.89	& 51.12	& 4.94	& 1.02     \T\B\\
\hline

\end{tabular}
\end{table*}

\begin{sidewaystable}
\centering
\caption{The characterised Step I beams.\label{tab:CharacterisedStep1Beams}}
\begin{tabular}{@{}ccccccccccc}
\hline

 & \multicolumn{2}{c}{Beam} & $\langle p_{z}\rangle$ (MeV/$c$) & $\sigma_{pz}$  (MeV/$c$)& $\varepsilon_{x}$  ($\pi$\,mm-rad)& $\alpha_{x}$ & $\beta_{x} (m)$   & $\varepsilon_{y}$ ($\pi$\,mm-rad) & $\alpha_{y}$ & $\beta_{y}$ (m)  \T\B\\

\cline{2-3}

 & $\varepsilon_{N}$ & $p_{z}$ &  &  &  &  &  \T\B\\ 
\hline
\multirow{9}{*}{$\mu^{-}$} & \multirow{3}{*}{3} & 140 & 171.58$\pm$2.39 & 22.81$\pm$ 0.32 & 2.28$\pm$0.12&0.50$\pm$0.01 & 1.49$\pm$0.09 & 0.95$\pm$0.05 &-0.55$\pm$0.28 & 3.62$\pm$0.18\T\\
 
 & & 200 & 223.24$\pm$2.72 & 24.02$\pm$ 0.29 & 1.74$\pm$0.09&0.49$\pm$0.01 & 1.69$\pm$0.10 & 0.78$\pm$0.04 &-0.50$\pm$0.25 & 3.71$\pm$0.19 \\ 
  
 & & 240 & 260.55$\pm$3.24 & 24.49$\pm$ 0.30 & 1.49$\pm$0.08&0.49$\pm$0.01 & 1.80$\pm$0.10 & 0.75$\pm$0.04 &-0.41$\pm$0.21 & 3.65$\pm$0.18 \B\\



 & \multirow{3}{*}{6} & 140 & 176.43$\pm$2.27 & 22.83$\pm$ 0.29 & 2.17$\pm$0.12&0.52$\pm$0.01 & 1.57$\pm$0.09 & 0.96$\pm$0.05 &-0.54$\pm$0.28 & 3.64$\pm$0.18 \T\\  

 &   & 200 & 232.22$\pm$2.51 & 23.62$\pm$ 0.26 & 1.53$\pm$0.08&0.55$\pm$0.01 & 1.85$\pm$0.10  &  0.78$\pm$0.04 &-0.51$\pm$0.26 & 3.80$\pm$0.19 \\    

 &   & 240 & 270.96$\pm$3.65 & 24.53$\pm$ 0.33 & 1.51$\pm$0.08&0.48$\pm$0.01 & 1.80$\pm$0.10 & 0.73$\pm$0.04 &-0.39$\pm$0.20 & 3.51$\pm$0.18\B \\ 


 & \multirow{3}{*}{10} & 140 & 183.46$\pm$2.35 & 22.75$\pm$ 0.29 & 2.01$\pm$0.11&0.53$\pm$0.01 & 1.62$\pm$0.09 &  0.92$\pm$0.05 &-0.56$\pm$-0.29 & 3.68$\pm$0.18 \T\\ 
  
 &   & 200 & 247.23$\pm$3.56 & 24.20$\pm$ 0.35 & 1.23$\pm$0.07&0.59$\pm$0.01 & 2.22$\pm$0.13& 0.75$\pm$0.04 &-0.52$\pm$-0.27 & 3.81$\pm$0.19 \\ 
  
 &   & 240 & 281.89$\pm$3.65 & 25.28$\pm$ 0.33 & 1.65$\pm$0.09&0.56$\pm$0.01 & 1.82$\pm$0.10 & 0.64$\pm$0.03 &-0.39$\pm$0.20 & 3.40$\pm$0.17 \B\\ 

\hline

\multirow{8}{*}{$\mu^{+}$} & \multirow{2}{*}{3} & 200 & 222.69$\pm$2.40 & 26.49$\pm$ 0.29 & 1.98$\pm$0.11&0.49$\pm$0.01 & 1.58$\pm$0.09 &  0.83$\pm$0.04 &-0.40$\pm$0.20 & 3.44$\pm$0.17 \T\\ 

 & & 240 & 257.97$\pm$2.83 & 26.37$\pm$ 0.29 & 1.59$\pm$0.08&0.57$\pm$0.01 & 1.87$\pm$0.11  & 0.76$\pm$0.04 &-0.31$\pm$0.16 & 3.40$\pm$0.17 \B\\ 


 & \multirow{3}{*}{6} & 140 & 176.45$\pm$1.98 & 24.36$\pm$ 0.27 & 2.32$\pm$0.12&0.45$\pm$0.01 & 1.50$\pm$0.09 & 0.95$\pm$0.05 &-0.48$\pm$0.25 & 3.59$\pm$0.18\T \\   

 &   & 200 & 229.16$\pm$2.36 & 25.87$\pm$ 0.27 & 1.91$\pm$0.10&0.50$\pm$0.01 & 1.61$\pm$0.09 & 0.81$\pm$0.04 &-0.38$\pm$0.19 & 3.42$\pm$0.17\\ 
  
 &   & 240 & 267.65$\pm$2.85 & 25.79$\pm$ 0.28 & 1.69$\pm$0.09&0.54$\pm$0.01 & 1.76$\pm$0.10 & 0.76$\pm$0.04 &0.26$\pm$0.14 & 3.23$\pm$0.16 \B\\ 


 & \multirow{3}{*}{10} & 140 & 182.42$\pm$2.05 & 23.87$\pm$ 0.27 & 2.16$\pm$0.12&0.47$\pm$0.01 & 1.56$\pm$0.09 &  0.92$\pm$0.05 &-0.48$\pm$0.24 & 3.59$\pm$0.18 \T\\  

 &   & 200 & 243.39$\pm$2.65 & 26.77$\pm$ 0.29 & 1.66$\pm$0.09&0.51$\pm$0.01 & 1.78$\pm$0.10   & 0.78$\pm$0.04 &-0.38$\pm$0.19 & 3.37$\pm$0.17 \\    

 &   & 240 & 274.77$\pm$2.94 & 24.79$\pm$ 0.27 & 1.78$\pm$0.09&0.51$\pm$0.01 & 1.65$\pm$0.09 &  0.76$\pm$0.04 &-0.22$\pm$0.11 & 3.07$\pm$0.15\B\\ 
\hline
\end{tabular}

\end{sidewaystable}

The measured emittances and optical parameters are given in Table~\ref{tab:CharacterisedStep1Beams}. The horizontal and vertical beta functions lie in the ranges 1.49\,m$ < \beta_{x} < 2.22$\,m and 3.07\,m$ < \beta_y < 3.81$\,m.  The values of the horizontal and vertical $\alpha$ parameters, $0.45  < \alpha_x < 0.59$ and $-0.56  < \alpha_y < -0.22$, show that the beams converge to a horizontal focus roughly 700\,mm downstream of TOF1 but diverge vertically. The emittances will be increased by scattering in TOF1. 

The measured horizontal emittances and simulations agree to within 10\%.
Some of the emittance of the beams can be attributed to multiple scattering in TOF0. The emittance growth in $x$~($y$) is expected to be $ \Delta \varepsilon_{x,y}^2 = \sigma_{x,y}^2 \theta_{\rm ms}^2 $ where $$\theta_{\rm ms}^2 = (13.6{\rm \,MeV/}c)^2/(p^2 \beta^2) \Delta X/X_{0} $$ is the mean square scattering angle in the $\Delta X = 0.125 X_{0}$ of material in TOF0. For 200\,MeV/$c$ muons and $\sigma_{x} = 70$\,mm, $\Delta \varepsilon = 1.9\,\pi$\,mm-rad for a beam of zero divergence, although the effective emittance at TOF1 is limited by the aperture of the Q7--9 triplet. The fall in measured emittance with increasing $p_{z}$ seen in Figures~\ref{fig:Hemit} and \ref{fig:Vemit} can be attributed to scattering via the dependence of $\theta_{\rm ms}$ on $p_{z}$.

There is some emittance growth in the $\approx 8$\,m of air between TOF0 and TOF1. Since the Q7--9 triplet focusses horizontally but is weakly defocusing vertically, this emittance growth is less in the horizontal than the vertical plane. For an on-axis beam, $\delta \varepsilon_{y}$ is estimated to be less than $0.4\,\pi$\,mm-rad. The resolution correction described previously includes a small upwards correction for this emittance growth, and has the largest effect on the measured vertical emittance. The remaining disagreement between the measured and simulated vertical emittances can be attributed to the difference in RMS vertical beam size\footnote{The RMS beam size in Figure~\ref{fig:beamsizes} is calculated and shown without the $\chi^{2}<6$ cut to demonstrate the physical size of the beam, whereas the emittance calculation includes it.} shown in Figure~\ref{fig:beamsizes}.

\begin{figure}[h,t,b,p]
\centering
\includegraphics[scale=0.5]{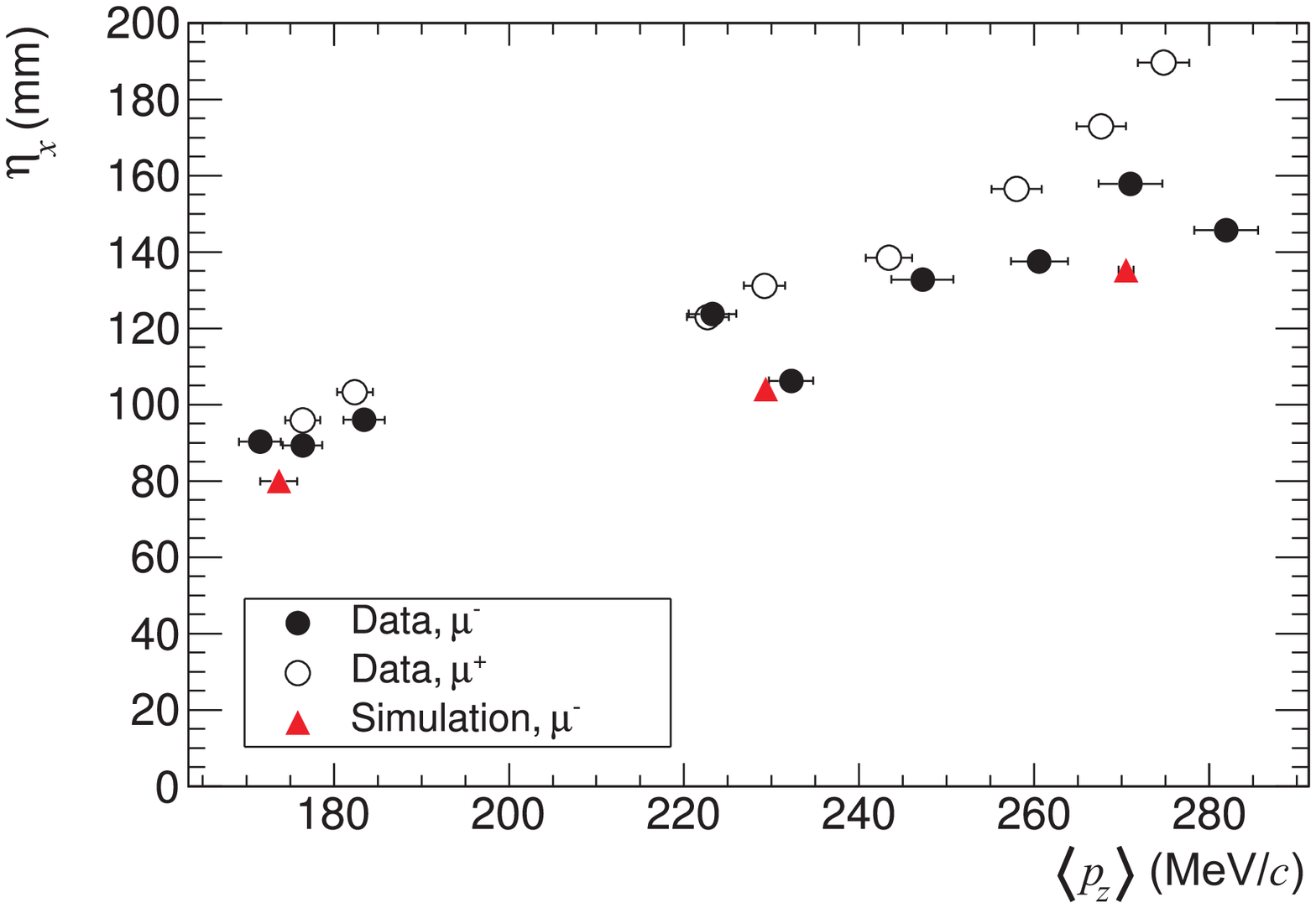}

\caption{The horizontal dispersion coefficient, $\eta$, versus mean $p_{z}$ for the seventeen beams.
Solid black circles: $\mu^{-}$ data, open black circles: $\mu^{+}$ data,
Solid red triangles:  $\mu^{-}$ simulation smeared with measurement resolution. The nominal  ``$p_{z}=140$''\,MeV/$c$ beams correspond to momenta in the range 170--190\,MeV/$c$, ``$p_{z}=200$'' to 220--250\,MeV/$c$, and ``$p_{z}=240$'' to 250--290\,MeV/$c$.
\label{fig:eta}}
\end{figure}

The measured horizontal emittances shown in Figure~\ref{fig:Hemit} include (for both data and simulation) the effect of dispersion. The dispersion in $x$ at the exit of the D2 bending magnet is transformed by the optics of the beam transport into dispersion in $x$ and $x'$ at the TOF1 measurement plane. The intrinsic horizontal emittances of the beams have been obtained from the covariance matrices by subtracting the dispersion characterised by $\eta$ and $\eta'$ \cite{etaref}:
\begin{eqnarray} 
\Sigma_{11} & \rightarrow \Sigma_{11} - \eta^2 \delta^2 \nonumber \\
\Sigma_{12} & \rightarrow \Sigma_{12} - \eta \eta' \delta^2 \nonumber \\
\Sigma_{11} & \rightarrow \Sigma_{11} - \eta^{'2} \delta^2 \nonumber 
\end{eqnarray} 
where $\eta = \langle x \delta\rangle/\langle\delta^2\rangle$, 
$\eta' = \langle x' \delta \rangle / \langle \delta^2\rangle$ and 
$\delta = (p_{z} - \bar{p}_{z})/\bar{p}_{z}$. 
Figure~\ref{fig:eta} shows $\eta$ versus $\langle p_{z} \rangle$ for all the beams and the simulations for the three negative beams. The dispersions are similar for the $\mu^{+}$ and $\mu^{-}$ beams and are reproduced by the simulations for the negative beams. The positive beam simulations are not shown as they did not reproduce the data well. The reason for this is under investigation. The dispersion-corrected intrinsic horizontal emittances and $\eta$ and $\eta'$ are given in Table~\ref{tab:IntrinsicStep1Beams}. The
intrinsic horizontal emittances are, on average, $0.25\,\pi$\,mm-rad smaller than the effective horizontal emittances.

\begin{table}
\centering
\caption{Horizontal dispersion and the intrinsic emittances of the Step I beams.\label{tab:IntrinsicStep1Beams}}

\begin{tabular}{@{}cccccccc}
\hline

 & \multicolumn{2}{c}{Beam} & $\eta_{x}$ (mm) & $\eta'_{x}$ (rad)  & $\varepsilon_{x}$ ($\pi$\,mm-rad) & $\alpha_{x}$ & $\beta_{x}$ (m) \T\B\\

\cline{2-3}

 & $\varepsilon_{N}$ & $p_{z}$ &   &   &  &  &  \T\B \\ 
\hline

\multirow{9}{*}{$\mu^{-}$} & \multirow{3}{*}{3} & 140 & 90.28 & 0.07 & 2.08$\pm$0.11 & 0.60$\pm$0.01 & 1.56$\pm$0.09\T\\
 
 & & 200 & 123.78 & 0.09 & 1.53$\pm$0.08 & 0.65$\pm$0.01 & 1.82$\pm$0.10\\ 
  
 & & 240 & 137.58 & 0.11 & 1.26$\pm$0.07 & 0.68$\pm$0.01 & 1.99$\pm$0.11\B\\ 


 & \multirow{3}{*}{6} & 140 & 89.37 & 0.08 & 1.97$\pm$0.11 & 0.64$\pm$0.01 & 1.66$\pm$0.09\T\\  

 &   & 200 &  106.27 & 0.10 & 1.31$\pm$0.07 & 0.72$\pm$0.01 & 2.06$\pm$0.12\\   

 &   & 240 & 157.91 & 0.11 & 1.26$\pm$0.07 & 0.68$\pm$0.01 & 1.98$\pm$0.11\B\\ 


 & \multirow{3}{*}{10} & 140 &  96.03 & 0.07 & 1.83$\pm$0.10 & 0.64$\pm$0.01 & 1.71$\pm$0.10\T\\ 
  
 &   & 200 &  132.78 & 0.08 & 1.04$\pm$0.06 & 0.79$\pm$0.01 & 2.47$\pm$0.14\\ 
  
 &   & 240 & 145.71 & 0.11 & 1.40$\pm$0.08 & 0.75$\pm$0.01 & 2.02$\pm$0.12\B\\ 

\hline

\multirow{8}{*}{$\mu^{+}$} & \multirow{2}{*}{3} & 200 & 122.96 & 0.03 & 1.85$\pm$0.10 & 0.56$\pm$0.00 & 1.58$\pm$0.09\T\\ 

 & & 240 &  156.47 & 0.03 & 1.45$\pm$0.08 & 0.66$\pm$0.01 & 1.87$\pm$0.11\B\\ 


 & \multirow{3}{*}{6} & 140 & 95.91 & 0.04 & 2.18$\pm$0.12 & 0.52$\pm$0.00 & 1.51$\pm$0.09\T\\   

 &   & 200 &  131.16 & 0.04 & 1.76$\pm$0.09 & 0.58$\pm$0.00 & 1.62$\pm$0.09\\ 
  
 &   & 240 &  172.97 & 0.04 & 1.54$\pm$0.08 & 0.64$\pm$0.01 & 1.76$\pm$0.10\B\\ 


 & \multirow{3}{*}{10} & 140 &  103.27 & 0.04 & 2.03$\pm$0.11 & 0.54$\pm$0.01 & 1.57$\pm$0.09\T\\  

 &   & 200 &  138.50 & 0.03 & 1.53$\pm$0.08 & 0.59$\pm$0.01 & 1.78$\pm$0.10\\   

 &   & 240 & 189.64 & 0.04 & 1.61$\pm$0.09 & 0.61$\pm$0.01 & 1.64$\pm$0.09\B\\ 

\hline
\end{tabular} 

\end{table}

\section{Summary}

A single-particle method for measuring the properties of the muon beams to be used by MICE has been developed. Timing measurements using two time-of-flight counters allow the momentum of single muons to be measured with a resolution of better than 4\,MeV/$c$ and a systematic error of $<3$\,MeV/$c$.  The ability to measure $p_{z}$ to this precision will complement the momentum measurements of the solenoidal spectrometers. For low transverse amplitude particles, the measurement of $p_{z}$ in the TOF counters is expected to have better resolution than that of the spectrometers, which are primarily designed for measuring $p_{t}$.

The same method allows the trace-space distributions at the entrance to MICE to be measured to $\approx 5\%$ and hence the emittances and dispersions of the beams. The emittances are found to be approximately 1.2--2.3\,$\pi$\,mm-rad horizontally and 0.6--1.0\,$\pi$\,mm-rad vertically; the average horizontal dispersion, $\eta$, is measured to be 129\,mm, although it depends on the nominal $(\varepsilon_{N}, p_{z})$ beam setting. The positive and negative muon beams are found to have very similar properties.

As a final check on the suitability of the beams for use by MICE, a set of measured muons  for
the (6,\,200) baseline beam was propagated from TOF1 to the diffuser and through a simulation of the experiment. 
Even without further software selection (for example, on the rather asymmetric momentum distribution)
the beam was found to be relatively well matched \cite{MRnuFact10}. In practice, some further 
fine-tuning of the magnet currents and diffuser thickness should be sufficient to generate a well-matched beam
suitable for the demonstration of ionisation cooling by MICE.  

\subsection*{Acknowledgements}

The work described here was made possible by grants from Department of Energy and National Science Foundation  (USA), the Instituto Nazionale di Fisica Nucleare (Italy), the Science and Technology Facilities Council (UK), the European Community under the European Commission Framework Programme 7, the Japan Society for the Promotion of Science and the Swiss National Science Foundation, in the framework of the SCOPES programme, whose support we gratefully acknowledge. We are also grateful to the staff of ISIS for the reliable operation of ISIS.

\bibliographystyle{spphys} 

\bibliography{references_v2}{}

\end{document}